\documentclass[reprint,superscriptaddress,amsmath,amssymb,aps,prx]{revtex4-1}

\usepackage{amsmath}
\usepackage{amsthm}
\usepackage[colorlinks=true,urlcolor=blue,citecolor=blue,linkcolor=blue]{hyperref}
\usepackage[shortlabels]{enumitem}
\usepackage{amsfonts}
\usepackage{graphicx}
\usepackage{bbold}
\usepackage{stmaryrd}
\usepackage{color}
\usepackage{hyperref}
\usepackage{verbatim}
\usepackage{cases}
\usepackage{amsfonts}
\usepackage{amssymb}
\usepackage[]{mathrsfs}
\usepackage{xcolor}
\usepackage{epsfig}
\usepackage{bm}
\usepackage{setspace}
\usepackage{enumerate}
\usepackage{dcolumn}
\usepackage{bm}
\usepackage{enumitem}

\usepackage{graphicx}
\usepackage{dcolumn}
\usepackage{bm}
\usepackage{color}
\usepackage{amssymb}
\usepackage{lipsum}
\usepackage{amsmath}
\usepackage{graphicx}
\usepackage{array}
\usepackage{multirow}

\usepackage[normalem]{ulem}

\begin{document}

\preprint{APS/123-QED}

\title{Topological Protection in non-Hermitian Haldane Honeycomb Lattices}

\author{Pablo Res\'endiz-V\'azquez}
\email{pablo.resendiz@correo.nucleares.unam.mx }
\affiliation{Instituto de Ciencias Nucleares, Universidad Nacional Aut\'onoma de
M\'exico, Apartado Postal 70-543, 04510 Cd. Mx., M\'exico}

\author{Konrad Tschernig}
\affiliation{Max-Born-Institut, Max-Born-Stra{\ss}e 2A, 12489 Berlin, Germany}
\affiliation{Humboldt-Universit\"at zu Berlin, Institut f\"ur Physik, AG Theoretische Optik \& Photonik, D-12489 Berlin, Germany}

\author{Armando~Perez-Leija}
\affiliation{Max-Born-Institut, Max-Born-Stra{\ss}e 2A, 12489 Berlin, Germany}
\affiliation{Humboldt-Universit\"at zu Berlin, Institut f\"ur Physik, AG Theoretische Optik \& Photonik, D-12489 Berlin, Germany}

\author{Kurt Busch}
\affiliation{Max-Born-Institut, Max-Born-Stra{\ss}e 2A, 12489 Berlin, Germany}
\affiliation{Humboldt-Universit\"at zu Berlin, Institut f\"ur Physik, AG Theoretische Optik \& Photonik, D-12489 Berlin, Germany}

\author{Roberto de J. Le\'on-Montiel}
\email{roberto.leon@nucleares.unam.mx}
\affiliation{Instituto de Ciencias Nucleares, Universidad Nacional Aut\'onoma de M\'exico, Apartado Postal 70-543, 04510 Cd. Mx., M\'exico}

\date{\today}

\begin{abstract}

Topological phenomena in non-Hermitian systems have recently become a subject of great interest in the photonics and condensed-matter communities. In particular, the possibility of observing topologically-protected edge states in non-Hermitian lattices has sparked an intensive search for systems where this kind of states are sustained. Here, we present the first study on the emergence of topological edge states in two-dimensional Haldane lattices exhibiting balanced gain and loss. In line with recent studies on other Chern insulator models, we show that edge states can be observed in the so-called broken $\mathcal{P}\mathcal{T}$-symmetric phase, that is, when the spectrum of the gain-loss-balanced system's Hamiltonian is not entirely real. More importantly, we find that such topologically protected edge states emerge irrespective of the lattice boundaries, namely zigzag, bearded or armchair.

\end{abstract}

\pacs{Valid PACS appear here}
\maketitle



\section{Introduction}

Over the last years, topological phenomena have attracted a tremendous interest in a wide variety of disciplines, including condensed-matter physics \citep{hasan2010,qi2011}, photonics \citep{Plotnik2014,Kitagawa2010,Lindner2011,Fang2012,Harari2018,Kremer2018, Weimann2017,Wang2019,Ozawa2019}, Floquet systems \citep{Torres2019}, ultracold atomic gases \citep{Aidelsburger2015,Stuhl2015,Mancini2015,Goldman2016}, acoustics \citep{Yang2015}, electronics \citep{Lee2018},  topoelectronics \citep{Lee2018,Ezawa2019}, and even chemistry \citep{Yuen2014}. Among different models where topological phenomena have been predicted and observed, the Haldane honeycomb lattice constitutes a paradigmatic example of a Hermitian system featuring a topological phase transition \cite{Haldane1988}. Indeed, the Haldane model represents a unique system where the quantum Hall effect \citep{Klitzing1980} is contained as an intrinsic lattice-band-structure property, rather than an external effect due to the presence of a strong magnetic field \citep{Chang2013}. Even though it was originally believed impossible to be implemented experimentally \citep{Haldane1988}, the Haldane model has been fundamental in the understanding of topological insulating (and conducting) phases and, more importantly, it has been the test bed for the experimental demonstration of topological edge-state protection in periodically modulated Floquet systems \citep{Jotzu2014}, and ferromagnetic insulators \citep{Kim2017}.

Hitherto, topological effects have been mostly explored in Hermitian systems \citep{Peng2014,Hu2011,Yao2018,Jin2017}. Yet, there is a growing interest in analyzing topological structures in non-Hermitian systems \cite{bergholtz2020}, particularly in conditions where balanced gain and loss is introduced, that is, in $\mathcal{P}\mathcal{T}$-symmetric systems. As first demonstrated by Carl M. Bender and Stefan Boetcher \citep{Bender1998}, $\mathcal{P}\mathcal{T}$-symmetric systems constitute an important subset of open quantum and classical systems, whose corresponding Hamiltonians are invariant under the combined operation of space and time reflection. Notably, depending on the gain-loss rate, these systems may exhibit a purely real or partially complex spectrum. When the former is observed, it is said that the system has an unbroken $\mathcal{P}\mathcal{T}$-symmetry; whereas when the spectrum is completely (or partially) complex, the system is said to be $\mathcal{P}\mathcal{T}$-symmetry broken \citep{Chen2017,roberto2018,Quiroz2019,Ozdemir2019,El2019}.

Quite recently, it has been shown that topological phase transitions may occur in non-Hermitian $\mathcal{P}\mathcal{T}$-symmetric photonic systems \citep{Xiao2017,Kremer2018,Zeuner2015}, as well as in photonic honeycomb lattices with armchair terminations \citep{Harari2015}. In particular, a recent work by Xiao and co-workers \citep{Xiao2019} has demonstrated the existence of topologically-protected edge states in one-dimensional $\mathcal{P}\mathcal{T}$-symmetry broken photonic networks, thus showing that $\mathcal{P}\mathcal{T}$ symmetry is not an essential condition for the observation of one-dimensional edge states.

Notably, the existence of topological protection in non-Hermitian systems has led to the creation of a new research line which focuses on the development of so-called topological lasers \citep{Bandres2018exp,Harari2018,Kawabata2019}. In the light of these findings, and since standard topological invariants---such as the Chern-number of the momentum-bulk Hamiltonian---may fail to correctly predict the existence of topological edge states in non-Hermitian systems \citep{Lee2016,bergholtz2020,Xiong2018,Ghatak2019,Helbig2019,Hofmann2019,Xiao2019obs}, many efforts are being devoted to investigate the benefits of the interplay between topology and $\mathcal{P}\mathcal{T}$ symmetry \citep{Alvarez2018,Ni2018,Poli2015,Schomerus2013,Yuce2018}. Prominently, there is an ongoing quest to generalize the bulk-boundary correspondence for non-Hermitian systems \citep{Kunst2018,Gong2018,Song2019}, which has revealed new phenomena exclusive to such non-Hermitian topological systems \citep{Alvarez2018non,Yoshida2019}.

In the present work we show that topological protected states can also be found in a two dimensional finite lattice, in contrast with \citep{Gong2018} and as it was predicted in \citep{Shen2018,Leykam2017}. As prototype system we use a Haldane topological lattice with balanced gain and loss, and show that edge states can be observed even when the spectrum of the system's Hamiltonian is not entirely real. Furthermore, we find that this behavior is universal in the sense that any geometry of the lattice edge, namely zigzag, bearded or armchair supports topological protection. This result contrasts with previous findings, where the observation of edge states in $\mathcal{P}\mathcal{T}$-symmetric hexagonal lattices was conditioned to armchair edges \citep{Harari2015,Leykam2017}. Our findings thus help enlightening the role of gain and loss in two-dimensional topological phenomena.


\begin{figure*}[t!]
\hspace*{1mm}
\includegraphics[width=\textwidth]{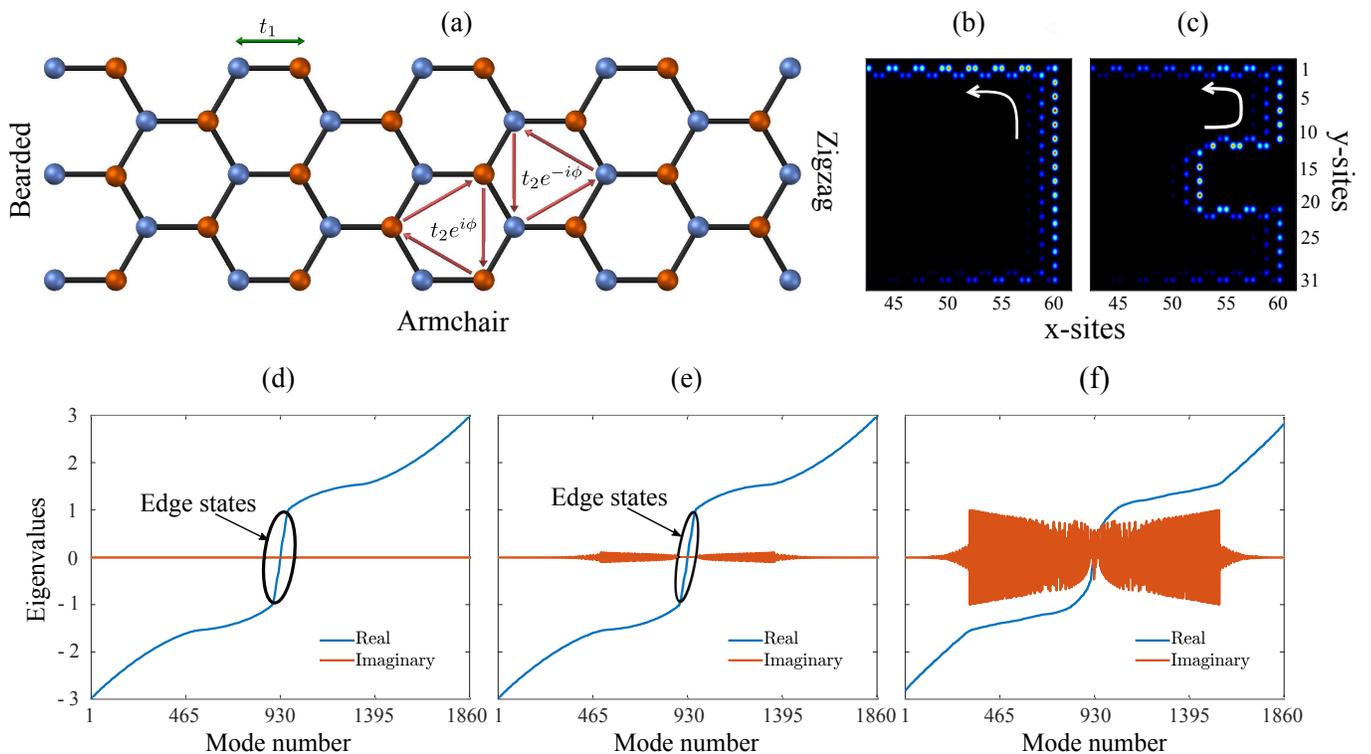}
\caption{(a) Schematic representation of the non-Hermitian honeycomb Haldane lattice with zigzag, bearded and armchair edges. The blue (orange) dots represent the amplifying (lossy) sites in the lattice. Note that the Haldane model includes two types of interactions between sites, a nearest-neighbor interaction $t_{1}$, shown as green arrows, and second-nearest-neighbor couplings $t_{2}$, with a constant phase $e^{\pm i\phi}$ for clockwise (anticlockwise) coupling direction, depicted by the red arrows. In order to implement balanced gain and loss, we set the on-site energies of the blue sites to $i \Gamma$ and those of the orange sites to $-i\Gamma$. (b) and (c) show the time evolution of an edge state over a $60 \times 31$ sites Haldane-ribbon lattice for $t_{1}=1.0$ s$^{-1}$, $t_{2}=0.3$ s$^{-1}$ and $\phi=\pi/2$; (b) shows the free evolution of the edge state, whereas (c) shows its propagation in the presence of a rectangular defect on the right edge of the lattice. (d), (e) and (f) show the real (blue line) and imaginary (orange line) energy eigenvalues per lattice mode for $\Gamma= 0$ s$^{-1}$, $\Gamma=0.1$ s$^{-1}$, and $\Gamma=1.0$ s$^{-1}$, respectively. The eigenvalues related to the topologically-protected edge states are shown in the region encircled by the black ellipse. Note that in (f) $\Gamma$ exceeds the critical gain-loss ratio $\Gamma_{c}=0.7$ s$^{-1}$, so no purely real eigenvalues are observed.}\label{fig:1}
\end{figure*}


\section{The model}

To study the emergence of edge states in non-Hermitian Haldane two-dimensional finite lattices (also known as Haldane ribbons), we consider the honeycomb lattice shown in Figure 1(a). As originally described by Haldane \citep{Haldane1988}, the dynamics of a single excitation in this type of lattice is described by the Hamiltonian
%
%
%
%
%
%
\begin{equation}
\hat{H}= \hat{H}_{1}+ \hat{H}_{2}+\hat{H}_{3},
\label{eq:hamiltonian}
\end{equation}
where the various $\hat{H}_{n}$ (with $n=1,2,3$) contributions describe the energy features of each lattice subunits (or sites), as well as the interaction between them. In particular
\begin{equation}
\hat{H}_{1} \equiv t_{1}\hspace{-1mm}\sum_{\langle n,m\rangle}\hat{c}^{\dagger}_{n}\hat{c}_{m}
\end{equation}
describes the nearest-neighbor interaction, with $\langle,\rangle$ denoting the summation over the nearest neighbors and $t_{1}$ being the coupling coefficient between them. The excitation creation and annihilation operators are denoted by $\hat{c}^{\dagger}_{n}$ and $\hat{c}_{n}$, respectively. Note that this term is needed in the Hamiltonian in order to obtain the so-called Dirac points and thus break the Inversion (IS) and Time Reversal Symmetry (TRS), which ultimately leads to the generation of topologically-protected edge states \citep{Lu2014}.

Furthermore, to gap out the Dirac cones that we created with $\hat{H}_{1}$, adding a second-nearest-neighbor complex coupling through the Hamiltonian $\hat{H}_{2}$ is required \citep{Haldane1988}
\begin{equation}
\hat{H}_{2} \equiv t_{2}\hspace{-3mm}\sum_{\langle \langle n,m \rangle \rangle} \hspace{-2mm} e^{\pm i\phi }\hat{c}^{\dagger}_{n}\hat{c}_{m},
\end{equation}
where the phase $\phi$ is defined along the arrows, being positive (negative) for clockwise (anticlockwise) coupling, as depicted in Fig. \ref{fig:1}(a). Note that the coupling $t_2$ denotes the interaction coefficient between second nearest neighbors, and so $\langle \langle,\rangle \rangle$ stands for summation over them. Interestingly, this term breaks the TRS as $\phi$ is changed; in particular, it allows the Hamiltonian to commute with the TRS operator, $\hat{T}$ for $\phi=\{0,\pi\}$; whereas TRS breaks for $\phi \neq \{0,\pi\}$.\\

Finally, in order to produce a topological phase transition, F.D.M. Haldane showed that breaking the IS is also required \citep{Haldane1988}. This can easily be done by adding an energy difference between sites. In the original Haldane model this was done by adding a real mass term $+M(-M)$ for odd(even) sites. Here, we break the IS by adding an imaginary balanced gain-loss parameter $+i\Gamma(-i\Gamma)$ in odd(even) sites with the term
\begin{equation}
\hat{H}_{3} \equiv i\Gamma\sum_{n \hspace{1mm} \text{odd} }  \hat{c}^{\dagger}_{n}\hat{c}_{n}-i\Gamma\sum_{n \hspace{1mm} \text{even}} \hat{c}^{\dagger}_{n}\hat{c}_{n}.
\end{equation}
A similar type of IS breaking has been employed in previous studies of non-Hermitian systems \citep{Klaiman2008,El2018,Szameit2011,Quiroz2019,Harari2015,Zhao1163,Yuce2019}, typically in the form of neighboring regions or strips of gain and loss. We want to emphasize, that in our approach gain and loss are dispersed over the lattice, such that each unit cell contains both. Yuce \emph{et al.} \citep{Yuce2019} investigated such an interspersed gain-loss distribution in the context of the 2D-Su-Schrieffer-Heeger model, but no real-valued edge-states were found. \\


\begin{figure}[t!]
\hspace*{-3mm}
\includegraphics[width=0.45\textwidth]{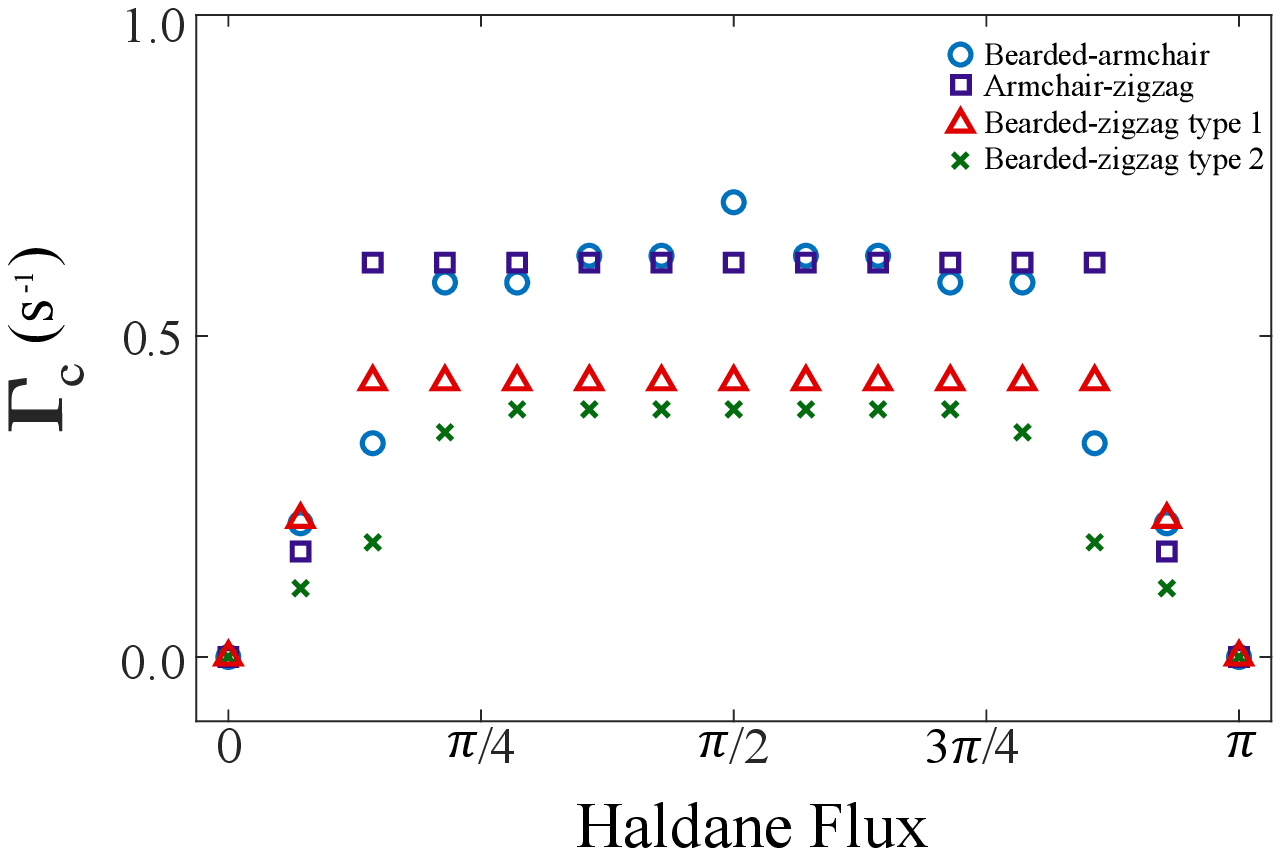}
\caption{Critical value of the gain-loss rate $\Gamma_{c}$ (defined as the maximum value of gain-loss for which at least twenty edge states remain within the dissipation/amplification-free region) as a function of the Haldane flux, $\phi$ for bearded-armchair (blue), armchair-zigzag (purple), bearded-zigzag type 1 (red) and bearded-zigzag type 2 (green). As discussed in the text, the topological phase transition occurs in a flux-range defined by $0<\phi<\pi$. Note that for the bearded-armchair termination, $\Gamma_{c}=0.70$ s$^{-1}$, while for armchair-zigzag termination, $\Gamma_{c}=0.61$ s$^{-1}$. For bearded-zigzag type 1 and 2, we obtain $\Gamma_{c}=0.38$ s$^{-1}$ and  $\Gamma_{c}= 0.42 $ s$^{-1}$, respectively.}\label{fig:2}
\end{figure}

\section{Results}

In the results that follow, we consider a finite rectangular Haldane ribbon comprising 1860 sites. Although we have explored different terminations for the lattice, here we present the results for a lattice with armchair-bearded termination. Similar results (that we provide in appendix A) can be found for zigzag-armchair and bearded-zigzag terminations, which implies that, in general, the edge geometry of the sample does not play a role in the observation of topological edge protection as long as the system is described by a ribbon. Interestingly, this contrasts with the findings of Harari and co-workers, where the observation of edge states in $\mathcal{P}\mathcal{T}$-symmetric hexagonal lattices is conditioned to armchair edges \citep{Harari2015}. The Hamiltonian parameters are set to $t_1 = 1\;\text{s}^{-1}$, $t_2 = 0.3\;\text{s}^{-1}$, $\phi = \pi /2$, and $\Gamma$ takes values within the range $0\leq\Gamma\leq 1$. It is worth pointing out that the location in the space of parameters $(\phi,t_{2})$ determines the topological gap proportional to $t_{2}\sin(\phi)$ \citep{Harari2018}. This is why we fix the value of the flux to be $\phi=\pi/2$, where the maximum energy band gap is reached for any value of $t_{2}$.

The band dispersion diagram for the Haldane model has an open band-gap with two edges traversing the bulk with opposite velocities, and these modes cannot be moved out from the gap by modifying the edge terminations \citep{Lu2014}. The time evolution of the system thus shows the topological protection of the edge-mode unidirectional propagation, which is caused by the TRS breaking and consequently the absence of counter-propagating modes at the same frequency as the edge modes, see Fig.~\ref{fig:1}(b). As the excitation can only move in one direction, the presence of a lattice defect does not affect its propagation and thus it travels arround the imperfection, as depicted in Fig.~\ref{fig:1}(c). It is important to remark that in order to find the proper initial condition for the observation of edge modes, we need to compute the spectrum of the Hamiltonian in Eq.~\eqref{eq:hamiltonian}, and identify the energies corresponding to the edge modes, which are located at the topological band gap between the two bulk energy states [top and bottom parts of the blue curve of Fig.~\ref{fig:1}(d)]. Note that in order to identify the protected edge states, we have followed the analysis presented by previous authors \citep{Bandres2016,Harari2015}, where the finiteness, or quasiperiodicity, of the lattice does not allow one to obtain a periodic band diagram. Finally, once we identified the edge-mode eigenfunctions, we generate the initial condition as a Gaussian distribution around the central eigenmode, which constitutes the proper state to observe topological edge protection.\\

One of the main goals of this work is to analyze the behavior of topological protection in the presence of non-Hermitian contributions. As we show next, there exists a critical gain-loss ratio $\Gamma_c$ below which we find completely real regions in the spectrum of the Hamiltonian in Eq.~\eqref{eq:hamiltonian}. Interestingly, this region contains the eigenvalues (and corresponding eigenvectors) that preserve topological protection [see Fig.~\ref{fig:1}(e)], and thus we can still find the proper initial condition for the generation (and preservation) of edge states even at the $\mathcal{P}\mathcal{T}$-broken phase. As one might expect, when increasing $\Gamma$ to larger values topological protection is lost, as all eigenvalues become complex, see Fig.~\ref{fig:1}(f). To be precise, we define $\Gamma_c$ as the maximum value of gain-loss for which at least twenty edge states remain within the dissipation/amplification-free region. It is important to remark that there is a close relationship between the system size (the number of sites) and the value of the critical gain-loss rate at which the system supports unidirectional edge states $\Gamma_{c}$: the larger the system, the larger value of $\Gamma_{c}$. In the model studied here, the critical value of loss-gain for the observation of topological protection is found to be $\Gamma_{c}= 0.7t_{1}$ for bearded-armchair termination and $\Gamma_{c}= 0.61t_{1}$ for armchair-zigzag termination, whereas $\Gamma_{c}= 0.38t_{1}$ and  $\Gamma_{c}= 0.42t_{1}$ for bearded-zigzag type 1 and 2 terminations (see Appendix A for details on the armchair-zigzag and bearded-zigzag terminations). Finally, we have explored the relation between $\Gamma_{c}$ and the Haldane flux, $\phi$. Figure (\ref{fig:2}) shows the results for Haldane ribbons with four different terminations. In the case of the bearded-armchair termination, it is shown that the $\Gamma_{c}$ reaches its maximum at $\phi=\pi /2$, while for the other cases, $\Gamma_{c}$ can be reached for some $\frac{\pi}{4}\leq\phi<\frac{3\pi}{4}$. Note that for all terminations there exists a region in the parameter-space $(\Gamma,\phi)$ - the area below the points plotted in Fig.~(\ref{fig:2}) - where protected edge-states are supported. This is particularly relevant for experimental realizations of the model, where a precise control of the Haldane flux might be cumbersome to reach.

It is an interesting matter to examine how the Chern-number of the Hermitian bulk Haldane model $\hat{H}_{1+2}(\vec{k})=d_0 \mathbb{1}+\vec{d}\cdot \vec{\hat{\sigma}}$ \citep{Asb_th_2016,bookLuo}, correctly predicts the existence of edge-states only for a finite range of the parameter $\Gamma$. To do so we first realize that, in Bloch-space, the non-Hermitian term $\hat{H}_3$ contributes as $\hat{H}_3(\vec{k})=i\Gamma \hat{\sigma}_z$. This manifests as a constant displacement of the torus into the complex plane $\vec{d}(k_x,k_y)\rightarrow \left(d_x(\vec{k}),d_y(\vec{k}),d_z(\vec{k})+i\Gamma\right)$. In fact, a formal redefinition of the origin $(0,0,0)\rightarrow(0,0,i\Gamma)$ readily shows that the Chern-number remains unchanged. However, as $\Gamma$ increases, it acts partially as a real mass-term on the bands $\varepsilon_\pm(\vec{k})=d_0 \pm \sqrt{d_x^2+d_y^2 +d_z^2 + 2id_z\Gamma -\Gamma^2}$, indicating that eventually they may touch and become degenerate. In this way, for sufficiently large $\Gamma$, the Chern-number loses its meaning and can no longer be used to predict the existence of edge-states.

\section{Conclusion}

In summary, we have shown the emergence of topological edge states in honeycomb two-dimensional lattices with balanced gain and loss. Surprisingly, we found that edge states can be observed even when $\mathcal{P}\mathcal{T}$ symmetry is broken. Furthermore, we have found that this behavior is universal in the sense that any geometry of the lattice edge, namely zigzag, bearded or armchair supports edge states. This contrasts with previous findings, where the observation of topological protection in hexagonal lattices was conditioned to armchair edges. Our results thus help elucidate the role of $\mathcal{P}\mathcal{T}$ symmetry in two-dimensional topological phenomena, and demonstrate that topological protection can exist in the archetypal Haldane model even in the presence of gain and loss.

\section{Acknowledgements}

This work was supported by CONACyT under the project CB-2016-01/284372 and by DGAPA-UNAM under the project PAPIIT-IN102920. P.R.-V. thanks CONACyT for financial support through the master's grant No. 895743. We thank Miguel A. Bandres for helpful discussions.

\vspace{5mm}

\appendix

\section{Topological-protection analysis in different lattice-edge terminations}

\begin{figure*}[t!]
\hspace*{0mm}
\includegraphics[width=\textwidth]{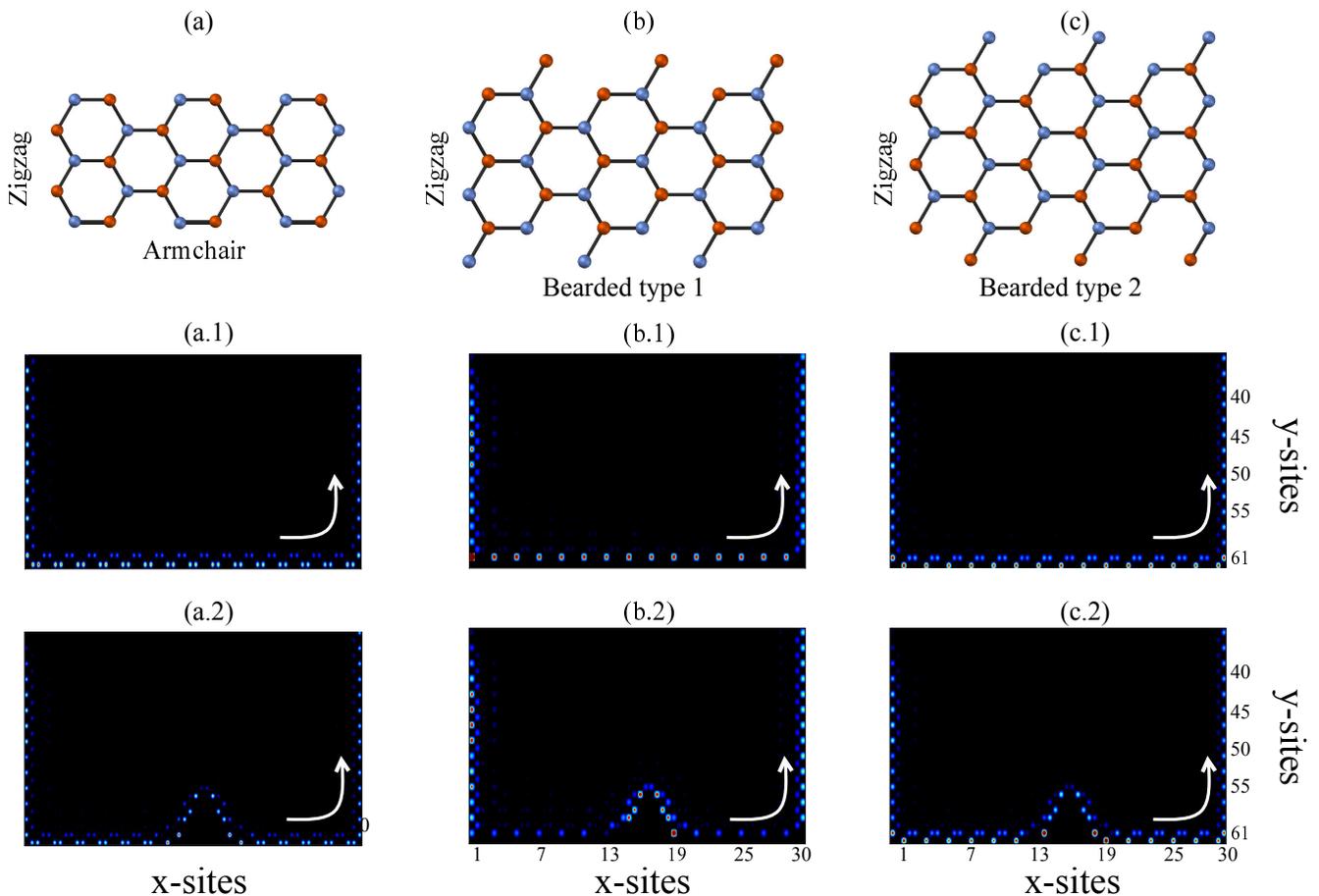}
\caption{(a)-(c) Schematic representation of the three geometries considered in this section. In (a.1)-(c.1) we show the free time evolution of an edge state over a $61 \times 30$ site Haldane-ribbon lattice with armchair-zigzag, bearded type 1-zigzag and bearded type 2-zigzag terminations for $t_{1}=1.0$ s$^{-1}$, $t_{2}=0.3$ s$^{-1}$ and $\phi=\pi/2$; whereas (a.2)-(c.2) show the propagation of the edge state in the presence of a finite triangular defect.
}\label{fig:3}
\end{figure*}

\begin{figure*}[t!]
\hspace*{0mm}
\includegraphics[width=\textwidth]{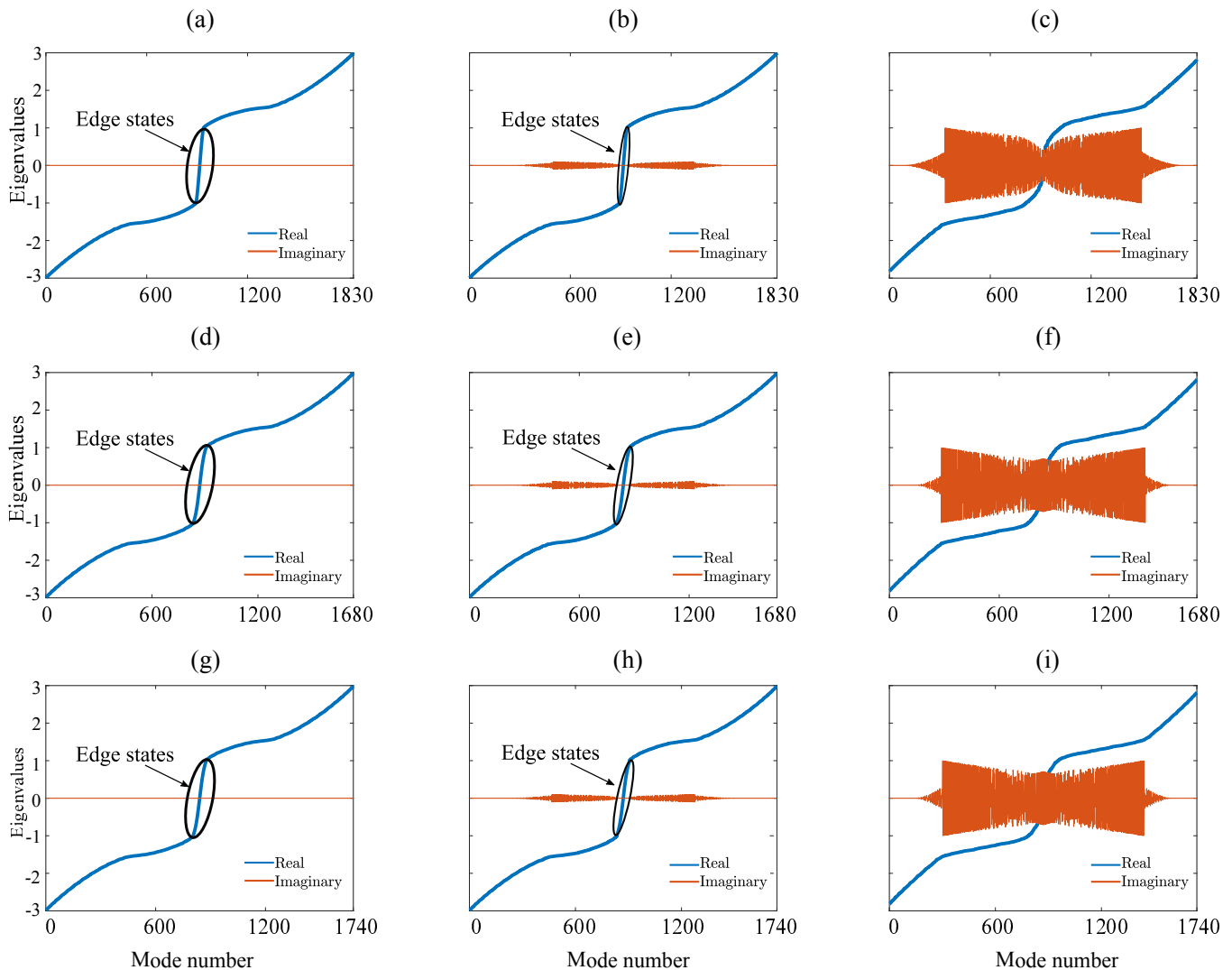}
\caption{ (a)-(i) Real (blue line) and imaginary (orange line) parts of the lattice eigenmodes for (a,d,g) $\Gamma=0$ s$^{-1}$, (b,e,h) $\Gamma=0.1$ s$^{-1}$ and (c,f,i) $\Gamma=1.0$ s$^{-1}$. (a)-(c) corresponds to armchair-zigzag terminations, whereas (d)-(f) and (g)-(i) correspond to bearded-zigzag type 1 and bearded-zigzag type 2 terminations, respectively. The eigenvalues related to the topologically-protected edge states are shown in the region encircled by the black ellipse. Note that in (c), (f) and (i) $\Gamma$ exceeds the critical gain-loss ratio for all the geometries, so purely real eigenvalues are completely absent.}\label{fig:4}
\end{figure*}

For the sake of completeness, we show in Figs.~(\ref{fig:3}) and~(\ref{fig:4}) the results for a Haldane-ribbon composed of $1830$, $1680$ and $1740$ sites with armchair-zigzag, and the two types of bearded-zigzag terminations [see Figs.~\ref{fig:3}(a)--(c)]. In the same fashion as in the bearded-armchair termination case, the band dispersion diagram of the Haldane model features edge modes connecting the bulk bands, thus it can support the back-scattering-free propagation, see Figs.~\ref{fig:3}(a.1)--(c.1). The edge excitation is not hampered by a defect of any shape or size, as depicted in Figs.~\ref{fig:3}(a.2), (b.2) and (c.2), now with triangular defects of $11\times 10$ sites. Finally, the eigenvalues of the Haldane-ribbon with these lattice terminations also show a region where they remain purely real, in these cases for $\Gamma \leq 0.61$  s$^{-1}$, $\Gamma \leq 0.38$  s$^{-1}$ and $\Gamma \leq 0.42$  s$^{-1}$ for armchair-zigzag, bearded-zigzag type 1 and bearded-zigzag type 2, respectively, see Fig.~\ref{fig:4}(b), (e) and (h). Beyond this value all the eigenvalues become complex and the topological protection is lost, as shown in Fig.~\ref{fig:4}(c), (f) and (i).

\vfill

\bibliography{biblio}

\end{document}